\documentclass[manuscript,screen]{acmart}

\AtBeginDocument{%
  \providecommand\BibTeX{{%
    \normalfont B\kern-0.5em{\scshape i\kern-0.25em b}\kern-0.8em\TeX}}}

\usepackage{hyperref}
\setcopyright{none}
\renewcommand\footnotetextcopyrightpermission[1]{}

\acmBooktitle{Woodstock '18: ACM Symposium on Neural Gaze Detection,
  June 03--05, 2018, Woodstock, NY}
\acmPrice{15.00}
\acmISBN{978-1-4503-XXXX-X/18/06}

\begin{document}

\title{MuseumViz - Towards Visualizing Online Museum Collections}

\author{Dheeraj Vagavolu}
\email{cs17b028@iittp.ac.in}
\author{Akhila Sri Manasa Venigalla} 
\email{cs19d504@iittp.ac.in}
\author{Sridhar Chimalakonda}
\email{ch@iittp.ac.in}
\affiliation{
\institution{\\
Research in Intelligent Software \& Human Analytics (RISHA) Lab,
\\Department of Computer Science and Engineering, \\
Indian Institute of Technology Tirupati}
 \city{Tirupati}
 \country{India.}
}

\begin{abstract}


Despite the growth of online museums for India's cultural heritage data, there is limited increase in terms of visitors. Over the years, online museums adopted many techniques to improve the overall user experience. However, many Indian online museums display artifacts as \textit{lists} and \textit{grids} with basic search functionality, making it less visually appealing and difficult to comprehend. Our work aims to enhance the user experience of accessing Indian online museums by utilizing advancements in information visualization. Hence, we propose \textit{MuseumViz}, a framework which processes data from online museums and visualizes it using four different interactive visualizations: the \textit{Network Graph}, \textit{TreepMap}, \textit{Polygon Chart} and \textit{SunBurst Chart}. We demonstrate \textit{MuseumViz} on a total of 723 cultural heritage artifacts present in the Archaeological Survey of India, Goa. Based on our evaluation with 25 users, about 83\% of them find it easier and more comprehensible to browse cultural heritage artifacts through \textit{MuseumViz}.

\end{abstract}



\maketitle

\section{Introduction}

In light of recent digitization, we have witnessed the growth of online museums that showcase the diverse cultural heritage collections taken from galleries, libraries, archives and museums \cite{windhager2018visualization}. By displaying the information about the cultural heritage artifacts, these online museums help promote these artifacts to a broader range of audiences \cite{styliani2009virtual}. Researchers have used various technical advances to improve the overall user experience, such as displaying cultural heritage information through mixed reality \cite{gheorghiu20143d} and 3D modelling \cite{thudt2012bohemian}. For instance, DIVE \cite{de2015dive} makes use of \textit{Linked Data} which allows users to search for cultural heritage artifacts through links between historical objects and events. Tietz et al. utilize knowledge graphs to deal with sparse meta-data in historic collections \cite{tietz2020knowledge}. 

In order to make online museums more interactive, researchers have used information visualization techniques to display the numerous cultural heritage artifacts in a more interactive and comprehensible way to the users  \cite{cecotti2020virtual}\cite{windhager2018visualization}. To visualize the extensive collections of cultural heritage artifacts, online museums have developed new types of web-based interfaces \cite{kalay2007new}\cite{parry2007recoding}. These interfaces leverage information visualization methods such as integrating Maps \cite{simon2016peripleo}, and slideshows \cite{crissaff2017aries} to enhance the user experience and allow for a better comprehension of the cultural heritage artifacts by the users \cite{windhager2018visualization}. The various visualization techniques enable the users to view the data with different perspectives \cite{sula2013quantifying}\cite{dork2017one}. For instance, \textit{Histograph} combines the timeline view with histograms targeting the relations between people, places and events \cite{novak2014histograph}. Some of the other temporal approaches include combining timelines with Ring charts \cite{deufemia2012investigative} and ontologies \cite{damiano2015visual}. Similar interfaces for non-temporal cultural heritage data utilize plots \cite{algee2012viewshare}, sets \cite{crockett2016direct}, and networks \cite{windhager2018visualization}.

In India, several steps towards digitization have been undertaken for the preservation of cultural heritage artifacts present in museums and galleries across the country \footnote{The National Portal of Museum for India: \url{http://museumsofindia.gov.in/repository/}}. The Digital Library of India initiative attempts to digitize one million textual artifacts, including but not limited to books, journals, and manuscripts and make them accessible to users all over the world \cite{joshi2006digital}. A similar attempt to digitize cultural heritage artifacts across 11 National Museums of India was done by Centre for Development of Advanced Computing, India (C-DAC)\footnote{Centre for Development of Advanced Computing, India: \url{https://www.cdac.in/index.aspx?id=mc_hc_jatan_virtual_museum}}. C-DAC developed the National Portal of Museum for India in which the cultural heritage artifacts are displayed in the form of grids and can be browsed through simple keyword searches. However, a limitation to this approach is that the user has to search and browse through a large number of catalogued cultural heritage artifacts that are often displayed as long lists and grids.

While there are several visualization solutions for museums across the globe, each of them is either region-specific, or target individual museums \cite{dork2017one} and are not available for Indian online museums. Even though it is possible to develop a tool for a single museum in India, it is effort-intensive to develop separate tools for 11 National Museums and multiple other museums \footnote{City wise museums in India: \url{https://www.museumsofindia.org/}}. Hence, there is a need to automate the process. We believe that the digitization of India's cultural heritage can be accelerated by utilizing the advances made in the representation of cultural heritage artifacts in the literature. For example, visualizing the relationship among different artifacts based on the textual information present in the Indian National Portal could help museum enthusiasts gain broader insights about multiple artifacts in the museum. Therefore, in this paper, we propose \textit{MuseumViz}, a framework that helps to visualize cultural heritage data in interactive ways to users and can be used for multiple Indian museums\footnote{MuseumViz demo: \url{https://museum-visualization.herokuapp.com}}. The contributions of this paper are:

\begin{itemize}
    \item \textbf{\textit{Visualization}} - \textit{MuseumViz} renders text-heavy cultural heritage data into four different visualizations: \textit{Network Graph}, \textit{TreeMap}, \textit{SunBurst Chart}, and \textit{Polygon Chart} focusing on different aspects of data such as connectivity and hierarchy.
    
    \item \textbf{\textit{Automation}} - The framework streamlines the process of collecting data from online museums and ultimately converting them into multiple data-driven visualizations for the users.

\end{itemize}

To demonstrate the framework, we automatically generate four different visualizations for the Archaeological Survey of India, Goa\footnote{Archaeological Survey of India, Goa: \url{http://museumsofindia.gov.in/repository/museum/gom\_goa}} which has a total of 723 cultural heritage artifacts. We show that the collections from it can be utilized to generate interactive visualizations for the users. We surveyed 25 users, asking them to compare their experience of using \textit{MuseumViz} against using the National Portal of India. The results show that by using \textit{MuseumViz}, 83\% of the users experienced a better overall user experience than using the existing web portal.

\section{Design and Development of \textit{MuseumViz}}
\label{dims}



\subsection{Design Decisions}

\textit{MuseumViz} is a framework that facilitates the interactive visualization of cultural heritage artifacts of existing collections taken from online museums and archives. \textit{MuseumViz} framework is modular, which allows easy modifications for targeting multiple online museums and allowing the integration of different types of visualizations.

We focus on multiple visualizations to allow the data to be comprehended in different perspectives. \textit{MuseumViz} renders the data in four different visualizations: the \textit{Network Graph}, \textit{TreepMap}, \textit{SunBurst Chart} and \textit{Polygon Chart}, each targeting different aspects of the cultural heritage data as shown in Table \ref{tab:vis_tab}. The interactive nature of these visualizations enables the users to navigate the artifacts based on different dimensions of the data with ease. \textit{Origin Place}, \textit{Object Type}, \textit{Dynasty} and \textit{Material} are the different dimensions that \textit{MuseumViz} draws from the meta-data.


\begin{table}[]
\caption{Four Types of Visualizations by \textit{MuseumViz}}

\label{tab:vis_tab}

\begin{tabular}{|l|l|}
\hline
\textbf{Visualization} & \textbf{Target aspect in the museum data}                                                 \\ \hline
Network Graph          & Focuses on connections between various artifacts present in the data                      \\ \hline
TreeMap                & Focuses on classification of cultural heritage artifacts                                  \\ \hline
SunBurst Chart         & Deals with the proportionality of the cultural heritage data based on the classifications \\ \hline
\end{tabular}
\end{table}





\begin{figure*}
    \centering
    \includegraphics[width=\textwidth]{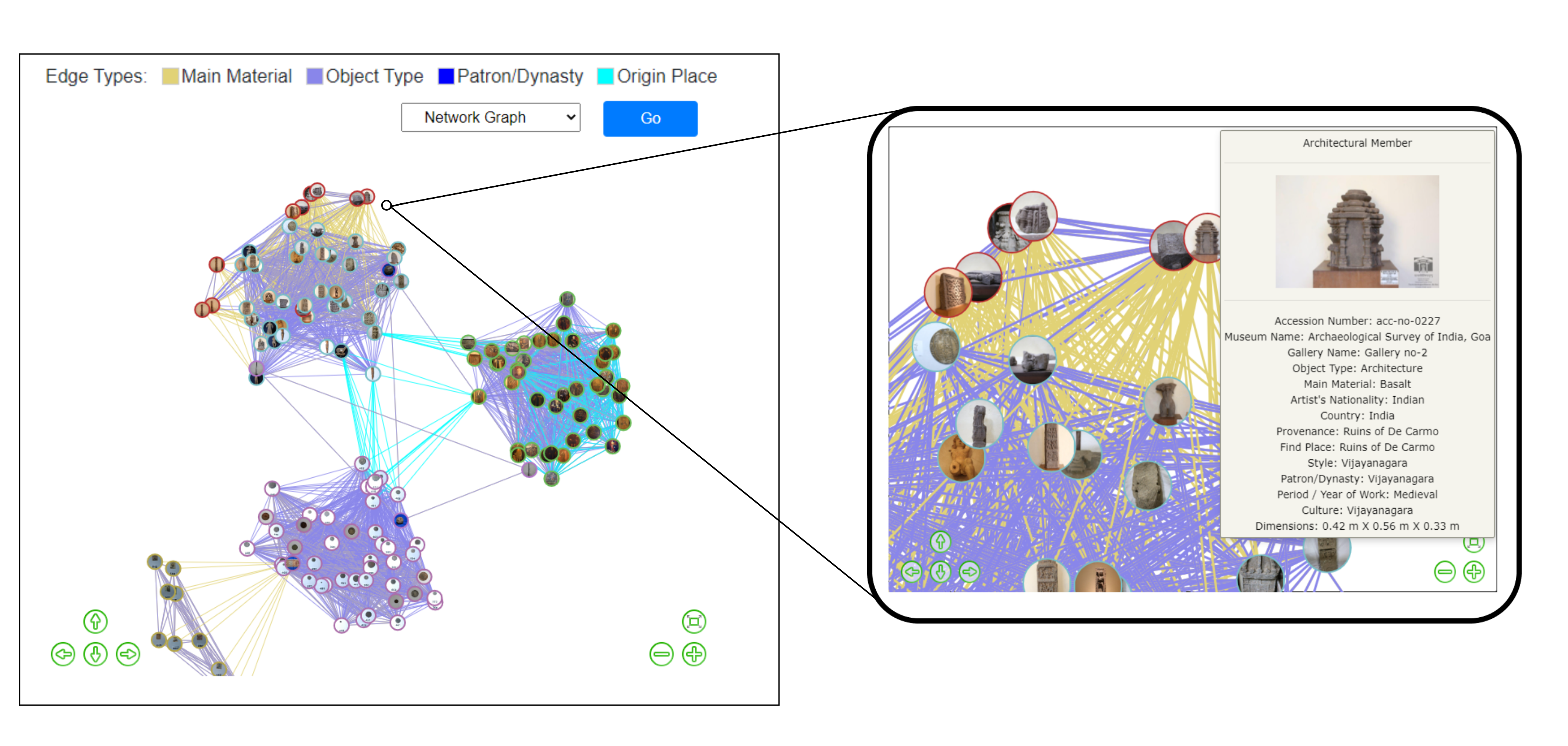}
    \caption{Network Graph of the artifacts from National Museum of Goa}
    \label{fig:net}
\end{figure*}    

\begin{figure*}
    \centering
    \includegraphics[width=\textwidth]{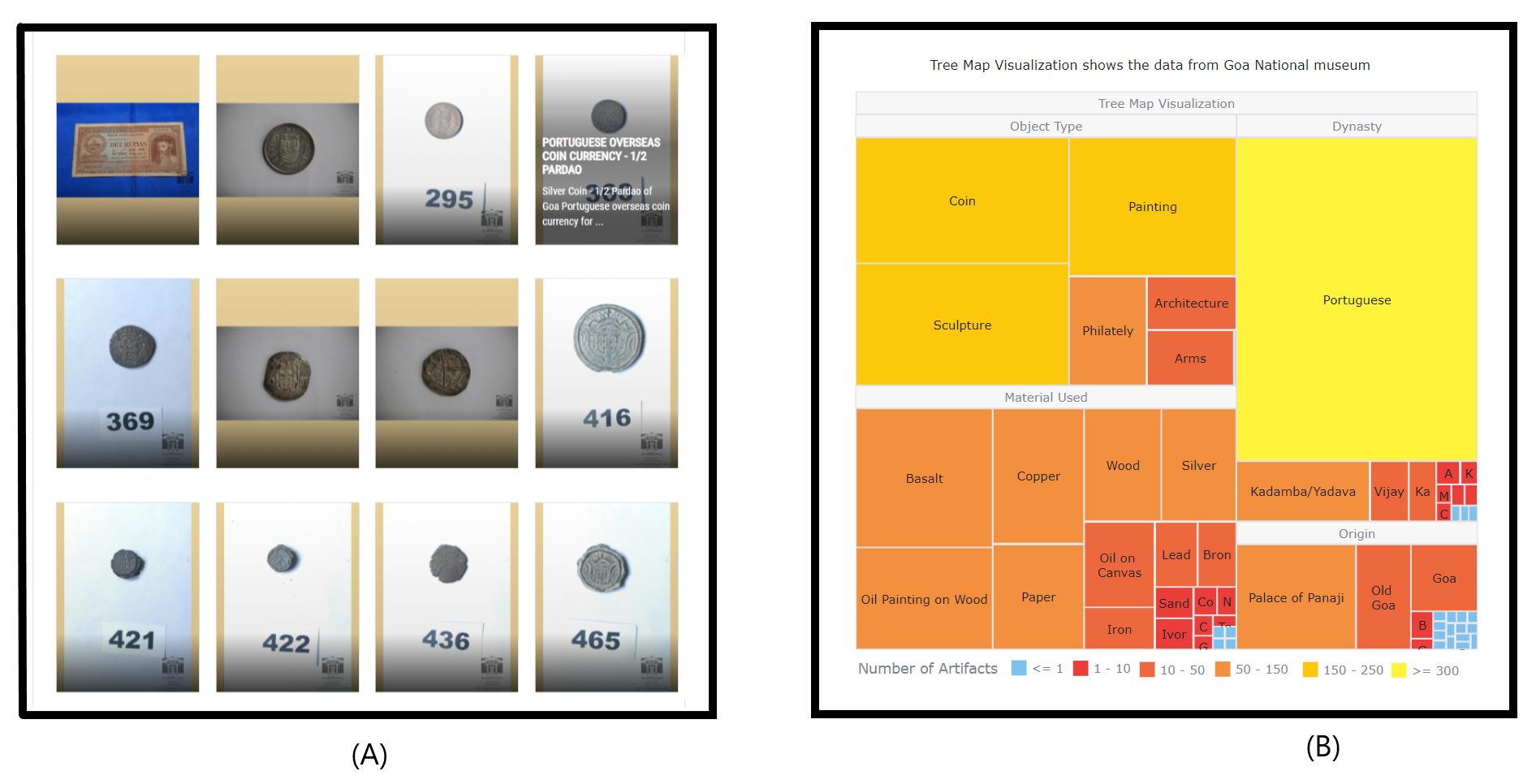}
    \caption{A) Interface of Archaeological Survey of India (Goa) B) TreeMap visualization of artifacts from Archaeological Survey of India (Goa) using \textit{MuseumViz}}
    \label{fig:compare}
\end{figure*}


\subsection{Development}

We developed the \textit{MuseumViz} framework as two different components that can generate interactive visualizations from online cultural heritage collections when invoked in a pipelined manner, as shown in Fig. \ref{fig:tool_arch}. \textit{Data Processor} and \textit{Visualizer} are the two main components of \textit{MuseumViz}. \textit{Data Processor} pertains to the collection and processing of data, whereas the \textit{Visualizer} component deals with displaying interactive browser-based visualizations.

\subsubsection{Data Processor}

The \textit{Data Processor} component is designed to collect data from existing online museums and then forward the processed data to the visualization module. It takes the target website and its blueprint and generates JSON files containing the data of all the cultural heritage artifacts. \textit{Data Processor} then processes the JSON files to convert them into input files specific to each visualization. The development of this module is done in \textit{python} and a combination of \textit{selenium}\footnote{Selenium library: \url{https://pypi.org/project/selenium/}} and \textit{BeautifulSoup4}\footnote{BeautifulSoup4 library: \url{https://pypi.org/project/beautifulsoup4/}} libraries. The current design of this component targets the National Portal of Museum for India developed by C-DAC. This component can be modified to target other online museums.


\subsubsection{Visualizer}
The \textit{Visualizer} component is developed as a standalone web application using \textit{Flask}\footnote{Flask framework: \url{https://pypi.org/project/Flask/}} and \textit{Javascript}\footnote{Javascript: \url{https://www.javascript.com/}}. \textit{Visualizer} takes the processed JSON files from the data processor and generates browser-based visualizations. The web interface renders the cultural heritage data into four different visualizations as mentioned in Table \ref{tab:vis_tab}, each showing various aspects of the data. The interface enables the users to switch between the available visualizations using a drop-down menu. The Network Graph is developed using \textit{Vis.js} \footnote{VisJs library: \url{https://visjs.org}}, a browser-based visualization library. The \textit{TreeMap}, \textit{SunBurst Chart} and \textit{Polygon Chart} are developed using \textit{Anychart}\footnote{Anychart library: \url{https://www.anychart.com}}, which is a lightweight browser based charting library. These visualizations are created as separate sub-modules and are highly modular. Hence the framework can be easily updated and modified to accommodate different visualizations using different existing APIs.


\section{User Scenario}

Consider the following situation, \textit{Moksha} is a university student who wants to write a report about the various cultural heritage artifacts present in the National Museum of Goa. \textit{Moksha} visits the National Portal of Museum for India by C-DAC to learn about the cultural artifacts hosted at the National Museum of Goa. Even though she can find all the artifacts in the portal, she finds it challenging to navigate various artifacts based on their meta-data. \textit{Moksha} comes across the \textit{MuseumViz's} visualization interface, which has already collected data about the various artifacts from the National Portal. At the landing page of the portal, \textit{Moksha} is prompted to select a specific visualization. She chooses the \textit{Network Graph} to understand the connections among different cultural heritage artifacts. \textit{Moksha} is now shown all the artifacts as nodes and different types of edges where each edge depicts a certain kind of connection among the artifacts based on the dimensions as illustrated in Fig. \ref{fig:net}. \textit{Moksha} now decides to learn about the hierarchy of artifacts based on different dimensions in the data. \textit{Moksha} now selects the \textit{TreeMap}, which shows different levels of classification of artifacts based on the meta-data, as depicted in Fig. \ref{fig:compare}. She can go deeper through the layers until she finds individual items based on the classification. Similarly, \textit{Moksha} uses the \textit{SunBurst Chart} and the \textit{Polygon Chart} to visualize and understand the distribution of data in different forms. 

\section{Evaluation and Results}

\begin{figure}
    \centering
    \includegraphics[width =\textwidth]{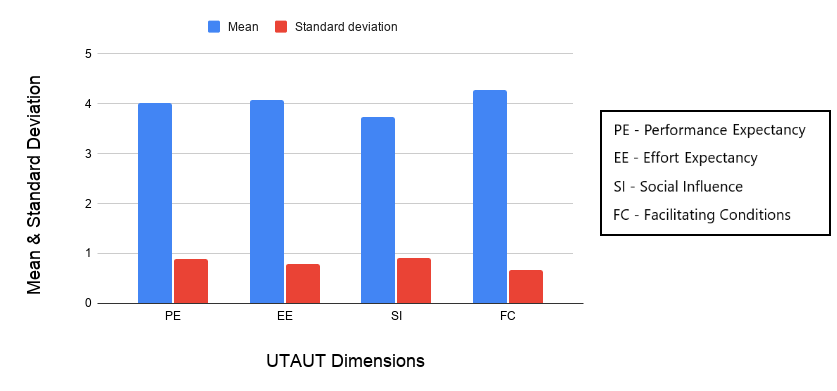}
    \caption{Results of user survey based on UTAUT model}
    \label{fig:results}
\end{figure}
The main aim of developing \textit{MuseumViz} is to improve the user experience for visitors of online museums by displaying various visualizations based on the relationships among multiple artifacts in the museum. Hence, we evaluate \textit{MuseumViz} to understand the usefulness and user experience, as perceived by the users. Models such as Technology Acceptance Model (TAM), Unified Theory of Acceptance and Use of Technology (UTAUT), Innovation Diffusion Theory (IDT) and so on support evaluation of a technological intervention through a questionnaire-based user survey that comprises of various dimensions which include usability, usefulness and ease of use \cite{venigalla2020dynamique, attuquayefio2014using}.
Existing technological innovations aimed towards improving user experience and usefulness have been evaluated using multiple adapted versions of these models, based on the nature of the innovation \cite{oshlyansky2007validating, attuquayefio2014using,  venigalla2020dynamique}. UTAUT model \cite{venkatesh2003user} deals with effort expectancy (EE), performance expectancy (PE), social influence (SI) and facilitating conditions (FC), which are influenced by gender, age, experience and voluntariness of use, and is also observed to be adapted for evaluation of technological interventions in museums \cite{kontiza2018museum, chen2012learning}. Considering the need for \textit{MuseumViz} to support a wide range of audience, irrespective of gender, age and experience, we observe the UTAUT model is a better fit for evaluating \textit{MuseumViz}. Hence, we conduct a user survey, using a 5-point Likert scale-based questionnaire designed to fit \textit{MuseumViz} evaluation aspects by adapting the UTAUT questionnaire. We also included other questions to understand the demographics of the users, such as age, gender and number of times they visited physical museums. We sent out the survey forms to 50 individuals with varied educational and professional backgrounds. 25 of these individuals have responded to the survey.

As we demonstrated the \textit{MuseumViz} framework for Archeological Survey of India, Goa\footnote{\url{http://museumsofindia.gov.in/repository/museum/gom\_goa}}, we requested all the participants of the survey to visit an online museum before responding to the survey. This is to ensure that the participants can distinguish between the visualizations being displayed by \textit{MuseumViz} from regular artifacts displayed by the museum. The questionnaire used and anonymized responses of the participants are presented here\footnote{\url{https://bit.ly/3sdgQhh}}. Based on the responses to the survey's demographic questions, we observe that 10 of the participants were female, and 15 of them were male. There were six participants in the age group 15-20, 15 participants in the age group 20-25, two participants with age group >35 and one participant each in age groups 25-30 and <15. Among the 25 participants, 18 participants have visited physical museums more than once, two have visited once, and 5 have never visited physical museums.   
The results of the user survey as presented in Fig. \ref{fig:results}, display higher mean values (>3.5), and lower standard deviation values (<1) for all dimensions, indicating that majority of the participants found \textit{MuseumViz} to be easy to use, useful and interesting, thus implying improved user experience. The mean 4.02 and standard deviation 0.88 for the performance expectancy (PE) dimension indicate that most participants had a common notion that \textit{MuseumViz} is useful and has increased their productivity.  \textit{MuseumViz} was observed to be easy to use, with the majority of the participants responding positively (mean = 4.06 and standard deviation = 0.78) to the questions based on effort expectancy (EE) dimension. Mean values for social influence (SI) and facilitating conditions (FC) dimensions are 3.74 and 4.27, respectively, with standard deviations equal to 0.91 and 0.66 for each of the dimensions. This indicates the willingness and possibility of participants to use \textit{MuseumViz}, which consequently indicates that \textit{MuseumViz} satisfies the goal of improving user experience to a considerable extent.

\section{Related Work}


Online museums facilitate the transmission of cultural heritage knowledge to a wide range of audience \cite{styliani2009virtual}\cite{aiello2019virtual}. However, studies also indicated that basic online websites might not attract more audience and suggest exploring solutions to make virtual museums more interactive \cite{fenu2018svevo}. One of the solutions suggested by Hinton et al. is to leverage techniques from the information visualization domain to make online museums interactive, and exploratory access to cultural heritage data \cite{whitelaw2010exploring}. Researchers have leveraged techniques such as storytelling/narration \cite{carrozzino2018comparing}\cite{swensen2019museums}, augmented reality \cite{venigalla2019towards}\cite{geronikolakis2020true}, and virtual reality \cite{cecotti2020virtual}\cite{jones2019reviewing} to make online museums interactive and more appealing to the users. Multiple visualization techniques such as usage of plots \cite{algee2012viewshare}, clusters and sets \cite{crockett2016direct}, geographical maps \cite{dumas2014artvis} and networks \cite{windhager2018visualization} have also been utilized in the development of online museums. 





In a survey by Windhager et al. \cite{windhager2018visualization}, around 80\% of the online museums utilize more than one visualization, as using a single visualization leads to over-complication of the visual representation of cultural heritage data \cite{mauri2013weaving}. To address this problem Mauri et al. \cite{mauri2013weaving} have developed a platform that integrates different types of entities and allows users to switch between different visualizations types coherently. Similarly, Kerracher et al. \cite{kerracher2014design} present a platform in which the user can select and switch between visualizations for different tasks such as browsing and searching of artifacts. 


These existing information visualization solutions are limited to individual museums or collection of museums over specific regions \cite{pescarin2014museums}\cite{terras2017accessing}; however, they are not available for India. For example, the National Portal of Museum for India introduces its cultural heritage through digital collections from 11 different National Museums across India. However, it uses a simple grid-based system to display the artifacts. The web portal focuses on showing the museum's objects as a list of categories and enables the searching of artifacts through keyword-based approach. The web portal also displays single object previews, which capture all the relevant textual metadata of a single artifact and displays it to the user.



Fig \ref{fig:compare}.A shows the interface of the National Portal of Museum for India. In contrast, by using multiple visualization techniques, the interface of \textit{MuseumViz} looks more appealing and informative as shown by the \textit{TreeMap} in the Fig \ref{fig:compare}.B. The \textit{TreeMap} presents the data in a hierarchical manner based on different dimensions extracted from the data. To this end, the goal of \textit{MuseumViz} is to act as a framework that can leverage multiple visualization techniques to improve the user experience and generalize this process for Indian online museums.



\section{Conclusion and Future Work}

Incorporating better visualization techniques in online museums provides the users with more insights about the cultural heritage data and improves their overall experience. The National Portal of Museum for India, developed by C-DAC, hosts cultural heritage artifacts from 11 different National Museums. However, it lacks the visual appeal that can help the users comprehend a large collection of cultural heritage artifacts. In this paper, we presented \textit{MuseumViz}, a framework specifically for enhancing Indian cultural heritage web portals using interactive visualizations. The framework uses four different  visualizations; the \textit{Network Graph}, the \textit{TreeMap}, the \textit{SunBurst Chart} and the \textit{Polygon Chart}, each targeting a different aspect of the cultural heritage data. We used the framework on the online museum of the Archaeological Survey of India, Goa, for demonstration. A total of 723 artifacts were collected from the portal and displayed using the four visualizations. The evaluation of the framework was done with a pool of 25 users through a remote survey. The results of the user survey with 25 users show that about 83\% of users prefer using \textit{MuseumViz} for browsing cultural heritage artifacts over the traditional web portal.

As part of the future work, we plan to incorporate different kinds of visualizations in the framework, providing additional perspectives to the cultural heritage data. The visualization interface is currently supported by desktop browsers. It can be further extended to support other devices such as mobile phones. We also plan to improve the framework to support a wide range of museum websites designed in different formats. This could be achieved by implementing appropriate machine learning and natural language processing-based approaches to detect and extract meta-data of the artifacts, irrespective of the structure of the website.  


\bibliographystyle{ACM-Reference-Format}
\bibliography{museumviz}

\end{document}